# An Open Distributed Architecture for Flexible Hybrid Assembly Systems: A Model Driven Engineering Approach

Kleanthis Thramboulidis
*Electrical and Computer Engineering Department, University of Patras, Greece.*

*Abstract*— **Assembly systems constitute one of the most important fields in today's industry. In this paper we propose an open distributed architecture for the engineering of evolvable flexible hybrid assembly systems. The proposed architecture is based on the model driven development paradigm. Models are used to represent structure and behavior and a domain specific engineering tool is defined to facilitate the assembly system engineer in the engineering process of the assembly system. Specific meta models are defined to capture domain knowledge to guide the engineer in the construction of the models required to construct the assembly system. This work is a specialization of our previous work that defined a SOA based framework for embedded industrial automation systems. It adapts and extends, in the assembly systems domain, the 3+1 SysML-view model architecture defined for the engineering of mechatronics Manufacturing systems. The proposed architecture can be used to develop a framework for evolvable flexible and reconfigurable assembly systems that would exploit the benefits the Cyber Physical paradigm utilizing web technologies, the IoT, the Cloud computing and Big Data.**

*Index Terms*—Assembly system Engineering, Evolvable assembly systems, SOA based Engineering tool, Meta modeling. Model driven engineering.

## I. Introduction

FOR a manufacturing product to be produced in its final form, its constituent parts should be properly assembled. This process is well known as *assembly process (AP)*. The entity that performs the assembly process is the *assembly system (AS)*. An AS is composed of the assembly process description that captures the logic of the assembly process and an underlying platform on which this logic is deployed and executed. The underlying platform is a mechatronic system and we refer to it with the term *Assembly System Platform (ASP)*.

Automation technology has considerably increased the effectiveness of *ASs*. However, as product variety increases due to the shift from mass production to mass customization, ASs must be designed and operated to handle such high variety. A review of state of the art research in the areas of AS design, planning and operations in the presence of product variety is given in [1]. Methods for assembly representation, sequence generation and assembly line balancing are reviewed and the authors conclude that many opportunities exist for future research in this domain. This is also claimed in [19], where authors admit that "there is still much to be done for making existing resources and processes more flexible, in an effort to satisfy the market demand and the increasing need for higher customisation on an individual basis." Among the challenges highlighted in [1] in the ASs domain we discriminate:

1. A new approach to provide manufacturing engineers with more comprehensive information with convenient data management features is required. The current assembly representations are considered limited in terms of the comprehensiveness of assembly information. Bill-of-Material (BOM) cannot directly represent the complex physical assembly processes and liaison graphs are considered as not suitable in representing hierarchical functional structures.
2. An assembly representation enabling interoperability across different locations and software platforms is required.
3. Standardization of assembly representation.
4. Determination of all possible assembly sequences as this greatly affects the total design process of a product.

This paper describes an open distributed architecture for flexible and evolvable *hybrid AS (HAS)*. We use the term hybrid to refer to ASs that include at least one human and one machine assembler. We argue that the proposed architecture addresses the above challenges and establishes a new approach for the engineering of systems in this domain. Moreover, this architecture facilitates the exploitation of technologies of Internet of Things (IoT), Cloud Computing, Big Data and Cyber Physical Systems in the domain of flexible and evolvable hybrid assembly systems.

The key concept of the proposed architecture is the use of meta models to formalize the domain knowledge and facilitate the job of *AP engineer*. Our approach does not consider BOM and liaison graphs as assembly representations, not even use this term. We discriminate between structural and behavioral information and capture this knowledge in three separate evolvable models, namely the product's structural model, the assembly process model and the assembly system platform model. The knowledge related with the construction of these models is captured in the corresponding meta models.

Ontologies may be utilized for a machine readable representation of the proposed framework and the Cloud could

K. Thramboulidis is with University of Patras, 26500 Greece. (e-mail: thrambo@ece.upatras.gr).

be used to establishing a distributed architecture for ASs. Web services are an effective communication mechanism to support a low coupling between the architecture elements. However, a more traditional distributed architecture based on technologies such as RPC or RMI may be utilized if the introduced high coupling between the elements of the architecture is not an issue.

Our objective is for the proposed architecture to have all the benefits of the automation technology preserving the flexibility of human based ASs. This leads to a new generation of automated hybrid ASs that are characterized by flexibility [21] and evolvability [4]. With the term flexible we mean an infrastructure that will enable the eight types of flexibility defined and described in [21][22]. Since, some researchers claim that "the acquisition of flexible technology as a direct response to changing markets is not necessarily the panacea it is widely believed to be" [22], it is a challenge to propose an approach that would limit the factors against the adoption of flexibility. Flexibility is discriminated from reconfigurability. As defined in [23], a manufacturing system is reconfigurable when it allows "cost-effective and rapid system changes, as needed and when needed, by incorporating principles of modularity, integrability, flexibility, scalability, convertibility, and diagnosability." On the other side flexibility is the characteristic of the system to provide generalized flexibility designed for the anticipated variations and built-in a priori. The proposed architecture provides the infrastructure required for both flexibility and reconfigurability.

The contribution of this paper is to define a set of meta-models and a reference open distributed architecture that exploits these meta models for the engineering of evolvable hybrid assembly systems. Meta models for the product, the assembly process and the assembly system platform are presented and their use in the development of flexible and evolvable hybrid ASs is described. The proposed approach is in the context of the Model Based Systems Engineering (MBSE) which is according to [25] "the formalized application of modeling to support system requirements, design, analysis, verification and validation activities beginning in the conceptual design phase and continuing throughout development and later life cycle phases."

The remainder of this paper is structured as following. In Section 2, background and related work is presented and discussed. Section 3 describes the proposed reference architecture and its key elements. Key concepts are identified and basic terminology is established. The proposed architecture for the *HAS Engineering Tool* is briefly presented. Section 4 presents the proposed in this work meta models for formalizing the knowledge required for the engineering process of evolvable hybrid assembly systems. The proposed meta models for product, assembly process and assembly system platform are presented and discussed. Finally, the paper is conclude in the last section.

## II. BACKGROUND AND RELATED WORK

The work presented in this paper is based on our previous work on Model Driven Engineering (MDE) of mechatronic systems, i.e., Model Integrated Mechatronics (MIM) [11] and 3+1 SysML-view model [12]. Initially, this work has been adapted for the modeling of Manufacturing Assembly Systems and then it was extended to address the specific challenges in this domain. Based on this, an open distributed architecture as a framework for the engineering of flexible assembly systems was defined. In this section we briefly refer to our previous work that is used as basis for the definition of the proposed open distributed architecture.

In our previous work [11] [20] we have adopted MDE and proposed Model Integrated Mechatronics, that is an approach for development of Mechatronic systems based on meta models. MDE [14] has been proved a successful paradigm in the development of information systems. It is proposed as a promising approach to alleviate the complexity of platforms and express domain concepts effectively [14]. MDE facilitates the construction of an integrated view of the system that is a prerequisite for identifying the optimal solution in the engineering process of complex systems. Other research groups have also reported on the successful use of this paradigm in mechatronic systems, e.g. [24]. MIM adopts MDE and proposes a framework for its application in the domain of mechatronics and in particular in Manufacturing systems engineering. The Mechatronic Component (MTC) is defined as the key construct in the development of Manufacturing systems. The Manufacturing system is defined as a composition of MTCs, which are integrated to collaborate with the objective to provide the system level behavior required to satisfy stakeholder needs. Those needs include among others, functional requirements, quality of service characteristics (QoS), and an optimal use of available resources. The MTC is defined as the element of the mechatronic system that is composed of a mechanical part, an electronic part, and a software part, which are tightly integrated and appear as an entity with well defined structure and behavior and is characterized by provided and required services offered through well defined interfaces. The MTC construct is ideal for the representation of assemblers as well as the other elements of the assembly platform.

The 3+1 SysML-view model is a realization of MIM based on the System Modeling Language (SysML). More specifically SysML is used in [12] as the primary artifact for modeling the mechatronic system in the mechatronic layer of the MIM Architecture and construct the core model of the whole system. SysML [5] has been proposed as an extension of the Unified Modeling Language (UML) [6], which is the defacto standard for the development of software systems. SysML is described as a general purpose graphical modeling language for specifying, designing, analyzing and verifying complex systems [25]. SysML is referred by INCOSE in [25] as one of the two explicit examples of systems modeling standards, the other being the ISO 10303-233 Application Protocol: Systems Engineering and Design (AP233), which are expected to have a significant impact on the application and use of Model-based systems engineering (MBSE).

The 3+1 SysML-view model and the corresponding extension of the V model, which adapts it to the specific needs of the mechatronics domain, provide a promising framework for the model based development of Manufacturing systems. In this paper we exploit the 3+1 SysML-view model for the modeling of assembly systems. An assembly system is defined in the Mechatronic layer as a composition of MTCs that represent the elements of the assembly system platform [20]. Next we build on the above work and define a framework for the engineering of flexible and/or evolvable assembly systems. A reference architecture is described exploiting SysML and the 3+1 SysML-view model to provide the infrastructure required for the development of this kind of systems. SysML has already been used by several research groups for the modeling of mechatronic systems, e.g., [26].

Service-oriented computing [15] provides the technologies required to address the need for distributed nature of assembly systems. Ontologies can be exploited to formalize the elements of the proposed infrastructure such as product, assembly process and assembly system platform, and make them machine-interpretable so that they can be more easily analyzed by the HAS Engineering Tool to assist the assembly system engineer in the decision making processes involved in the engineering process of hybrid assembly systems. The use of Ontologies and SOAs in assembly systems has already been reported by research groups, e.g., [27][28][29]. This paper does not explicitly refer to the exploitation of service oriented computing and ontologies in the presented framework. We have already presented in [13] a framework for exploiting these technologies in the engineering process of industrial automation systems. Encouraging results have also presented by other research groups.

In [8], authors describe an approach for evolvable assembly systems based on Ontologies. An assembly process ontology based on the Process Specification Language (PSL) [9] is presented. The process specification language (PSL) was defined according to [7] to be "a common language to all manufacturing applications, generic enough to be decoupled from any given application, and robust enough to be able to represent the necessary process information for any given application." To this direction it can be used for complete and correct exchange of process information among established applications. PSL does not provide assembly specific constructs and constraints thus the ontology presented in [8] is considered as an extension of PSL, which is considered as a solid basis for the definition of such an ontology. However, this ontology is based on the specific structure of the assembly process based on the constructs of task, operation and action. The task is used to define the sequence in which the components are being assembled to form the final product. The term Operation is used to define the steps required to put the components together including feeding, handling assembly, etc. The Actions construct is used to define the individual motions and other more hardware and control related activities. These three constructs correspond to three different level of the assembly process with the task assigned to the highest one and the action to the lowest one. We consider this approach unable to meet the requirements of our framework. It leads to platform dependent assembly process descriptions since the action refers to hardware and control related activities which are platform specific. Moreover, we disagree with the whole proposed structure for reasons we explain in this document. We consider as a prerequisite for a successful ontology a robust meta model. Thus, we first focus on the construction of meta models that will support the construction of platform independent assembly process descriptions.

Authors in [18] propose an assembly sequence generation algorithm based on CAD data. The paper also briefly presents the most important research works towards this direction. These works can be utilized in the context of the proposed framework to get the benefits of the meta model driven approach. The intention of this work is not to propose a new algorithm for assembly sequence generation but to automatically construct a model of the assembly process that would capture the dependencies between product components as they come from the product model. This model is independent of the assembly platform and will be the source from which the assembly platform specific assembly process will be generated taking into account the assembly platform model. This results to the definition of the assembly sequence based on the optimal utilization of the specific platform resources which are captured in the assembly platform model. This proposal is different from the above work in that it introduces a two step approach for the assembly sequence generation based on product and assembly platform models.

In [9] authors describe an approach called Assembly Process Micro-Planning which is based on PSL. The objective is to have the Computer Aided Assembly Process Planning tools and Computer Aided Assembly Process Simulation Tools to use the same assembly process description. They adopt four levels of Assembly process decomposition: Assembly Job (AJ), Assembly Task (AT), Assembly Operation (AO) and Assembly Action (AA). They use the construct of AJ to model behavior assigned to a worker or a team. The construct of AT is used to model assembly sequences in AJs that correspond to the behavior of assembling a part or a subassembly. The ATs of an AJ can be executed sequentially or in parallel. This approach is considered as assembly platform depended since from the first step of AJ definitions the assembly platform workers are considered.

Author in [30] describe a systematic approach based on process knowledge customization and meta models for manufacturing resources. Authors in [31] describe an integrated design model of assembly systems where decision regarding layout, assembly sequencing, task assignment and assemblers' locations are analyzed to obtain the optimal solution. To the best of our knowledge there is no open reference distributed architecture based on meta models for the engineering of evolvable and flexible hybrid assembly systems.

### III. A REFERENCE ARCHITECTURE FOR EVOLVABLE HYBRID ASSEMBLY SYSTEMS

An AS accepts as inputs the product's parts and their connectors (if any) and delivers as output the finished product as shown in Fig. 1.

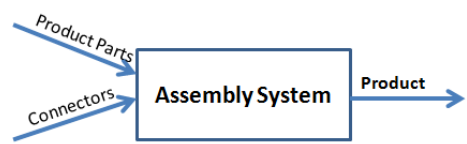

*Figure 1. The traditional assembly system has the assembly logic hardcoded in the hardware platform.*

Traditionally, assembly systems are constructed to perform a specific assembly process for a specific product. However, market demands for product variations and/or product enhancements impose the need for flexible ASs. The term flexible means that the system is able to change or to do different things. An AS is considered to be flexible if it is composed of a generic ASP that may accept as input also the assembly logic in terms of an assembly process description (as part of the assembly job), as shown in fig. 2. The assembly process description when deployed onto the ASP transforms it into a specific AS. We use the term *assembly job* to represent a job that is assigned to the assembly system. The assembly job contains information about the product to be produced, the number of items, the specific variation, quality parameters, etc.

ASs are mainly composed of assemblers and tools, both considered as resources of the assembly system platform. Assembler or assembly line worker is a human or machine (robot) that alone or in collaboration with other assemblers has the responsibility to take of the parts of products and put them together towards their production. A flexible assembly system platform provides the functionality of not only assigning assembly behavior to its assemblers but also specifying their collaboration.

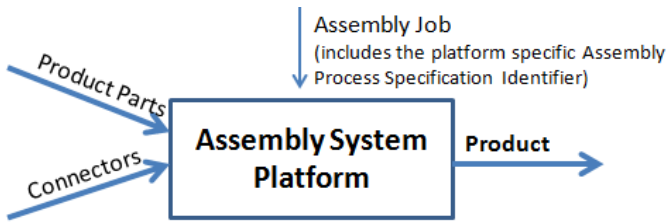

*Figure 2.* A flexible assembly system (Black box view).

*A. Using meta models for knowledge representation*

As claimed in [1], the design of an AS requires methods to represent the assembly components and hierarchy, and to generate the sequences of assembly. To this direction [1] reviews "the most commonly used assembly representation methods, including liaison and precedence graphs, and discuss how these methods are adopted for representation of products with variety." Bill-of-Material (BOM) and Liaison graphs are reported as commonly used assembly representation methods. In our approach we discriminate the structural from the behavioral information and we use two different models, i.e., the product's structural model and the behavioral model of the assembly system, called assembly process model, to capture this information. We capture the information regarding the product's structure, i.e., the product's parts and their physical contacts of joining, called liaisons, in what we call *product's structural model (PSM)*. Key concepts used for the construction of product's structural models are captured in the *product's structural meta model (PSMM)* to facilitate the assembly system engineer in the construction of robust and manageable structural models of products. BOM or liaison graphs could be used as input for the generation of the PSM which will be an instance of the PSMM. The proper mapping of the corresponding meta models, i.e., BOM and PSMM meta models, will enable an automatic model to model transformation for the automatic or semi automatic construction of PSMs from the BOMs.

We represent the behavior required by a system to perform the assembly of a product in what we call *assembly process model (APM)*. The challenge is to have an assembly system that would accept assembly requests and define dynamically its behavior utilizing the knowledge captured in the specific product's structural model, as shown in fig. 3.

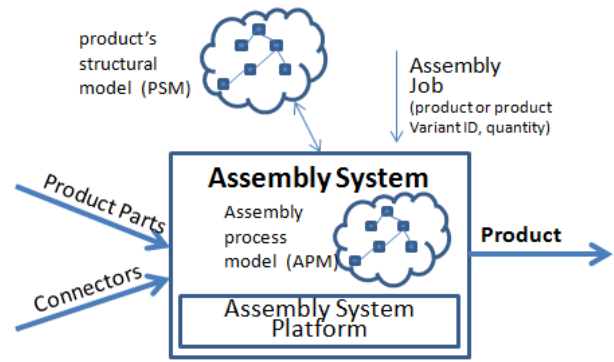

*Figure 3.* An evolvable assembly system accepts assembly requests in the form of assembly jobs and has the knowledge to be self transformed to an assembly system for performing the specific assembly job.

We formalize the knowledge related with the construction of AP models of in what we call *assembly process meta model (APMM)*. APMs are defined as instances of the APMM. We further discriminate between two *assembly process models (APMs)* defined by a two step procedure. The first model is the *platform independent APM* which captures the assembly logic required for the assembly of the final product in a platform independent way. It captures chunks of behavior that correspond to the assembly of product's parts without defining sequence or parallelization regarding their execution. The platform independent APM captures just the precedence constraints among the product's components as they are expressed in the corresponding product's structural model and those imposed by assembly constraints imposed by other factors, e.g., product's geometry. Thus *platform independent APM* captures all possible assembly scenarios of the specific product. The second model, the *platform specific APM*, describes the behavior that should be deployed to the ASP to transform it to an assembly system for the specific product. The platform specific APM is generated by refining the platform independent APM taking into account the structural and behavioral information that is captured in the model of the ASP. We formalize the knowledge related with the construction of *ASP models (ASPMs)* in what we call *assembly system platform meta model (ASPMM)*. ASPMs are defined as instances of the ASPMM.

*B. Properties of Assembly System*

The description of the AP should be in a readable and understandable by the system format for the AS to be able to realize it. For assembly systems composed of human assemblers, a human-readable assembly process description is required. Since humans may change their behavior based on different assembly process descriptions, by default the human based assembly system is flexible, evolvable and reconfigurable. Traditional automated assembly systems have lost this flexibility since they are traditionally constructed with the assembly process logic hardcoded (see figure 1). Any change in the assembly logic requires a reconstruction of the assembly system, that is a costly and time consuming process. Evolvable assembly systems [2-4] come to address this disadvantage of traditional automated assembly systems. According to [4], evolvability is defined as "the ability of complex systems to co-evolve with the changing requirements, to undergo modifications of different significance, from small adaptations on-the-fly to more important transformations." The term evolvability with the above definition overlaps with the term flexibility. We consider a system evolvable when it is able to change its behavior by its own to meet changing requirements as shown in figure 3. The framework described in this paper provides the infrastructure to support both flexibility and evolvability of the assembly system.

*C. The Hybrid Assembly System Engineering Tool*

For the above properties, i.e., flexibility and evolvability, to be realized and taking into account that an assembly system is a composition of possible heterogeneous assemblers, a uniform machine readable assembly process description is required. If we now take into account that human assemblers should participate in complex assembly systems then a uniform human and machine-readable assembly process specification is required. If this is not feasible automatic transformation between these representations should be provided. Moreover, these specifications should be easily deployed into the assembly system. A Engineering Tool, called *HAS Engineering Tool*, is proposed to address the need for constructing AP specifications and deploy these on hybrid ASs to perform the assembly process, as shown in Fig. 4.

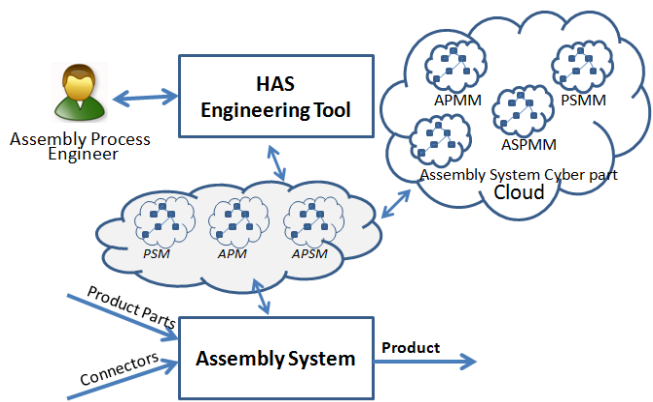

**Figure. 4.** *Using a HAS Engineering Tool for the configuration of Evolvable Hybrid Assembly Systems.*

The Assembly Process Engineer using the HAS Engineering Tool and utilizing the assembly system platform meta model (ASPMM) available in the Cloud, constructs the ASP model that represents the hybrid ASP that would be assigned the assembly job. The HAS Engineering Tool also allows the assembly process Engineer to construct the product's structural model (PSM) utilizing the corresponding meta model.

These models constitute along with the assembly process model the infrastructure that is required for generating and evaluating alternative assembly scenarios for the specific product. The specification of the selected as optimal assembly process is deployed on the hybrid ASP to construct the specialized AS required for the assembly of the specific product or product variation.

Thus, the objective of the HAS Engineering Tool is to support the engineer to construct the optimal AS that fulfils the requirements of the stakeholders for assembling the specific product or products or perform specific assembly job(s). More specifically the HAS Engineering Tool will allow the user:
a) to capture the logic of the assembly process,
b) to transform the assembly process logic into an assembly process specification for the specific platform,
c) analyze and evaluate alternative scenarios, and
d) deploy the so constructed assembly process specification on the assembly system platform.

Moreover, it will provide the functionality for modeling the assembly system platform. This model will be utilized during: a) the assembly logic description process, b) the transformation of the assembly job logic description into an assembly specification for the specific platform, and c) the deployment of this specification on the assembly system platform.

The HAS Engineering Tool may be developed with the traditional approach of developing engineering tools. In this case it is represented as a system, a is the case with figure 4. However, we propose a service oriented architecture development based on [13] where the infrastructure required to build a service-based Engineering Tool is presented.

In this case there is no one entity that constitutes the Engineering Tool but a set of distributed services that may interoperate to accomplish the engineering process. Services that would provide generic functionality as well as specific functionality required by the AP Engineer should be identified and properly defined. The definition of services in the assembly system domain is a challenge since it should be based on many parameters such as performance, flexibility, maintainability, and reuse.

Among these services we discriminate: a service to define and manage assembly Jobs, a service to define and manage product's structural models, a service to define and manage AP specifications, a service to deploy AP specifications, a service to define and manage ASP models, etc. For these services to interoperate in order to constitute a coherent HAS Engineering Tool, the constructed models are stored in the local Cloud and are identified using properly constructed URIs. Access to meta models on the Cloud is through URIs.

The assembly job contains the URI of the platform independent assembly process specification that captures the logic of performing one assembly of the final product.

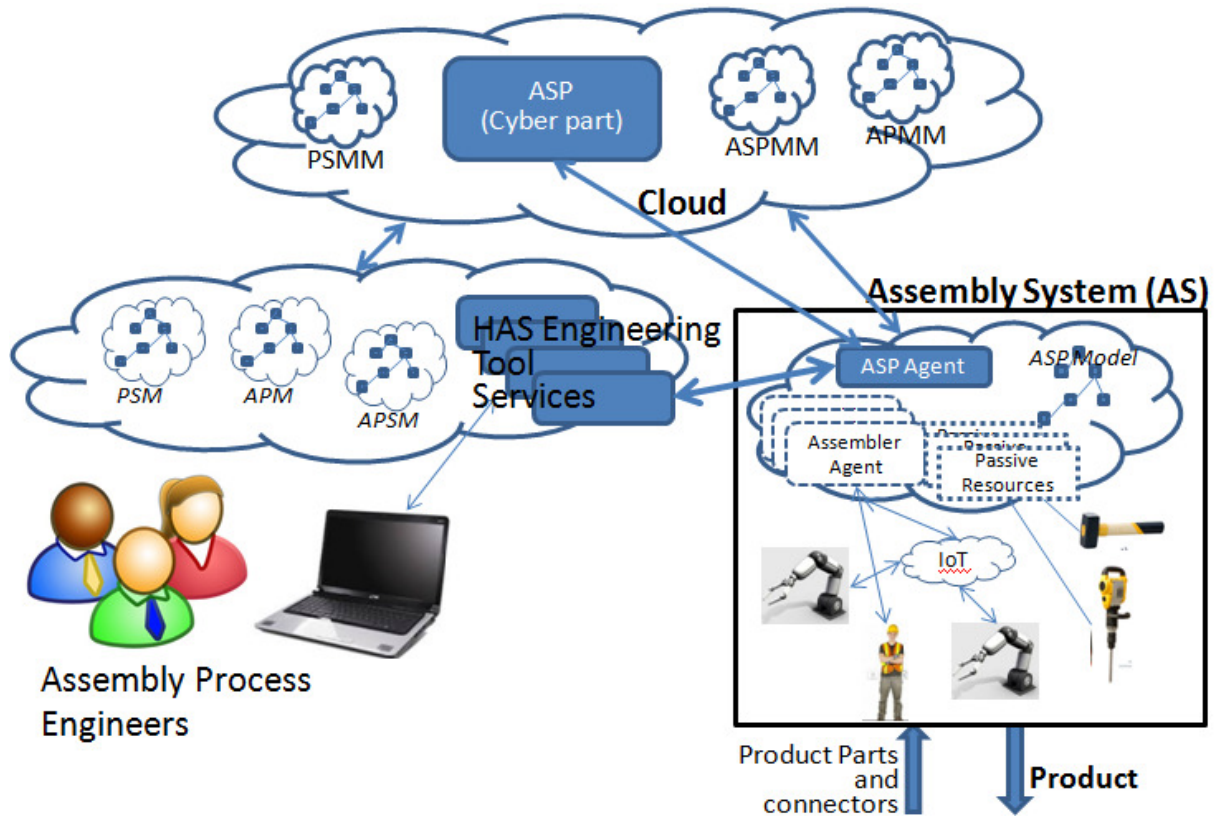

*Figure 5*. The proposed distributed Architecture for the Assembly System and the HAS Engineering Tool.

*D. The Architecture of the Assembly System*

Figure 5 captures also the core architecture of the Assembly System. Machine assemblers are interconnect through IoT while human assemblers are interconnected to the system platform through their agents in the local cloud. Passive resources also have their software representations that transforms these to smart resources. Assemblers, and other resources of the underlying infrastructure along with the software layer that represents the collaboration infrastructure knowledge constitute the ASP. Thus the ASP is a distributed entity and communicates with its actors through the ASP agent. The ASP agent transforms the physical infrastructure of the AS into a smart ASP. At the lower level the IoT already provides technologies for implementing a low an so flexible coupling between the ASP assemblers.

An interesting element of the proposed architecture is the Cyber part of the ASP that is shown in the Cloud captured in Figure 5. The ASP agent communicates with the Cloud Cyber part of the ASP to report data regarding the operation of the ASP. The Cloud Cyber part of the ASP collects data from all the instances of the specific ASP type all over the world and using Big Data technologies analyzes these data to produce knowledge regarding preventing maintenance, evolution, fault handling, etc. This knowledge is exploited by local ASP agents to transform every ASP instance into a self evolvable system. This Cloud Cyber part of the ASP if properly defined and maintained will greatly improve the system maintenance, operation, refinement, evolution, etc., all those benefits that emanate from the use of Cyber Physical concepts.

IV. THE PROPOSED META MODELS

The key concept of MDE is to consider models as the primary artifact in the Engineering process. For the models to be effectively constructed by the assembly process engineers a framework is required to capture the rules, constraints and capabilities of utilizing the available resources of the assembly system platform and guide to the optimal solution. We construct this infrastructure in terms of meta models. To this direction specific meta models are identified and developed to capture the domain logic regarding product assembly, assembly process and assembly system platform. The corresponding meta models are utilized by specific model editors to construct the models that will be used in the engineering of the assembly process.

Figure 6 captures the key concepts involved in the engineering process of assembly systems. As assembly job we consider an order that is given to an assembly system platform to produce specific number of a product. The assembly job contains the identifier of an assembly process of the specific product that will be realized by the assembly system platform to perform the assembly job. The assembly process specification is defined in terms of actions which are the smallest reusable units of behavior that can be used to describe assembly tasks. Actions are platform independent. For an assembly process specification to be deployed on an assembly system platform, the platform should have all the skills that realize the actions used by the APS.

A product may have several assembly processes each one describing a specific process of assembling the product's constituent parts. An assembly process is associated with a specific product or product variation and it may be realized by one or more assembly systems, as shown in figure 6. Key concept of the assembly process is the assembly process description which behavior that should exposed by the assembly system to accomplish the assembly of a product. An assembly process specification is constructed using constructs of higher layer of abstraction compared to action but this is not shown in this concept map. It is described in the assembly process meta model sub section.

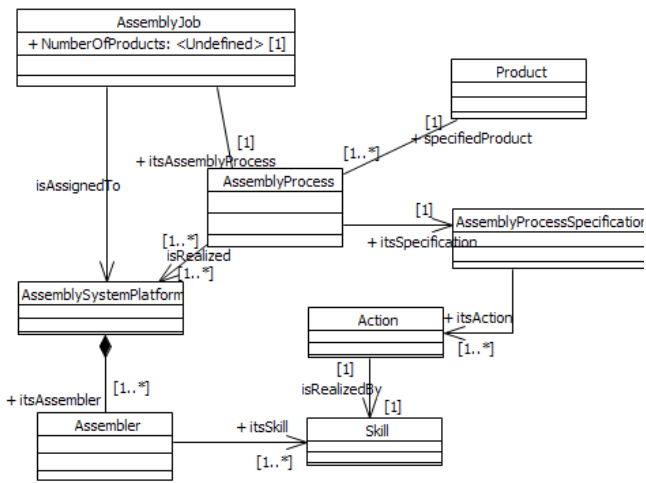

*Figure 6.* *The key concepts of the concept map.*

A. *The Assembly System Platform meta model*

For simple assembly tasks the assembly system may be composed of an assembler. However, as the complexity of the assembly process increases the assembly system is constructed as a composition of assembly sub systems each one performing a part of the assembly process, i.e., a sub assembly process, and their connectors, i.e., sub assembly system connectors. Fig. 7 presents the proposed assembly system platform meta model. An assembly platform sub system that may not be partitioned into lower layer assembly sub systems is called Assembler. The Assembler is the human or machine entity with the ability to perform assembly activities. Connectors interconnect assembly sub systems through their input and output ports. Connectors may be active, such as conveyors or passive. A passive connector is just as storage location that is considered as output port for the input assembly sub system of the connector and as input port for the output assembly sub system of the connector. The structure of the assembly system platform is captured in a model that is instance of the assembly system platform meta model.

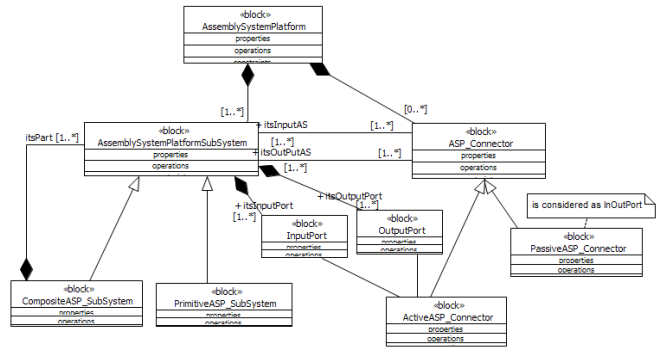

*Figure 7.* *The Assembly System Platform meta model (core part).*

B. *The Product's structure meta model*

The product model defines the constituent parts of a product and their interconnections using the proper connectors. The product model is a prerequisite for the definition of its assembly process. For the construction of the product model the product meta-model shown in Fig. 8 is used. Based on this a product is considered as a composition of sub assemblies (see SubAssembly block) and connectors (see Connector block). A sub assembly is either a primitive part, i.e., an element of the product that is not further decomposed into other constituent parts, or a Composite part, i.e., a system element that is a composition of other elements. A connector interconnects 2 or more SubAssemblies and is associated with their corresponding Liaison. A Liaison (ConnectionPort) is used to represent a connection point or a connection surface of the SubAssembly. The optional association between Product and SubAssembly and the one between Product and Connector are used to capture product variations.

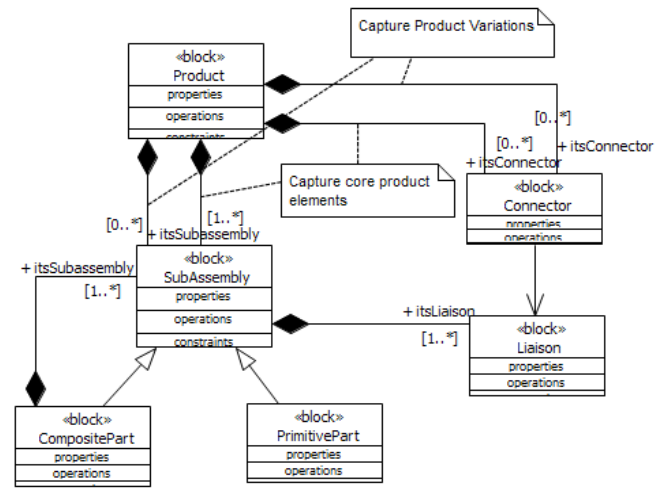

*Figure 8.* *The Product's structure meta model (core part).*

The assembly level of the composition tree in which every part belongs is captured in the SubAssembly or the CompositePart. It will be utilized to built the assembly process model.

## C. The Assembly Process meta model

In the assembly process meta model the assembly process is modeled as CompositionLevelAssemblyProcess. It represents in the model space the part of the assembly process that captures the assembly logic of the first level of decomposition of the product, i.e., composition level 0. This is shown in Fig. 10 as dcl-0. We define the aggregation hierarchy of a product as consisting of n levels of decomposition, which constitute the aggregation tree. The first level of decomposition, i.e., dcl-0, is the highest level of composition of the products aggregation tree. The dcl-n is the lowest level of the products aggregation tree. As shown in Fig. 9 a CompositionLevelAssemblyProcess is either a Primitive ChildsAssemblyProcess or a CompositeChildAssembly-Process. As PrimitiveChildsAssemblyProcess we represent in the modeling space the assembly process/sub-process that captures the assembly behavior that corresponds to the assembly of the composite part of a decomposition level in the case that all the constituent parts (child) of the composite part (parent) are primitive as defined in the product meta model. In the case that at least one of the child is composite, the parent's assembly process is modeled as CompositeChildAssembly Process. Each CompositionLevelAssemblyProcess is characterized by the level of decomposition of the products aggregation tree. Precedence constraints are captured at all levels of the hierarchical structure of the AP, as shown in figure 9. They are also captured at the Action level. It should be noted that the alternative to define the association between the Operation and Action as {ordered} was not adopted since the precedence constrain at this level allows also parallelism on action execution that leads to a better utilization of the ASP resources.

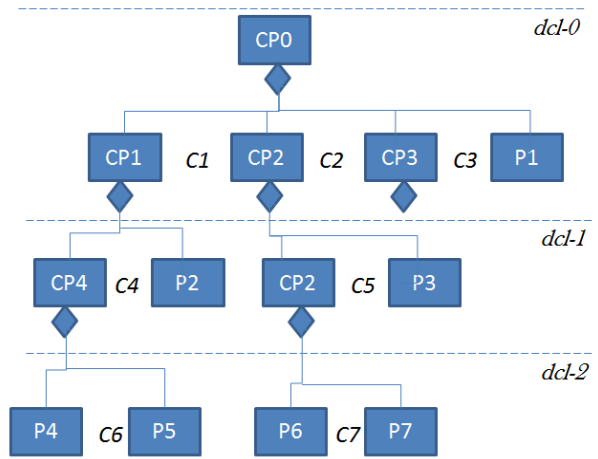

*Figure 10. Decomposition levels on a product's structural model.*

A CompositeChildAssemblyProcess is defined as a composition of the assembly processes of the composite childs of the corresponding decomposition level and the primitive assembly activities which are required for the assembly of the constituent parts of the parent at this decomposition level. A PrimitiveChildsAssemblyProcess is defined as a composition of primitive assembly activities (PrimitiveAssemblyActivity) which are required for the assembly of the constituent parts of the parent at this decomposition level.

A PrimitiveAssemblyActivity is used to represent in the model space the process of assembling two constituent parts of the n level. In case of complex parts with many liaisons more levels of PrimitiveAssemblyActivities may be defined to represent different levels of abstraction to handle the complexion in the assembly of the parts. It is advised to define one PrimitiveAssemblyActivity for any interconnection of two parts based on the corresponding Liaisons

This low level PrimitiveAssemblyActivity is described by the set of operations that should be executed in order to complete an interconnection based on the corresponding liaisons. Thus it is modeled as a composition of Operations. Among Operations we discriminate assembly operations, move operations handle operations, etc. An Operation is defined as a composition of actions. The Action is used to represent in the model space the primitive behavior of the assembly system. It has a well defined interface and semantic. Actions are predefined and standardized to facilitate a platform independent assembly process description. It is evident that for a platform independent assembly process description to be realizable on a given assembly platform the actions used for the assembly process description should be implemented in terms of skills by the assembly system platform workers.

For the product of figure 10 the AP is composed of one CompositeChildAssemblyProcesses and at least three PrimitiveAssemblyActivities one for each one of the interconations shown in figure as C1, C2 and C3 for the level dcl-0.

The proposed assembly process meta model is based on four levels of process decomposition, i.e., Process, activity,

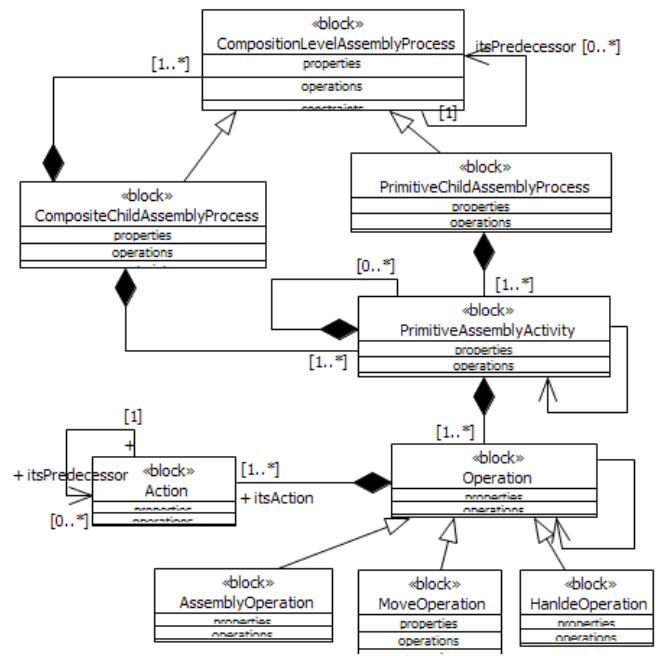

*Figure 9. The Assembly Process meta model (core part).*

operation and action. Its structure is based on the product aggregation tree that is considered as an instance of the product meta model. This allows the assembly process to automatically embed in its structure the composition constraints imposed by the product model. Moreover, it allows a platform independent description that is the prerequisite for the effective application of the MDE paradigm in the assembly system domain.

## V. Concluding Remarks

An open distributed architecture for the model driven engineering of assembly systems has been described. Meta models were used to capture the domain knowledge and are used to facilitate the engineer in the specification of the assembly process in its final executable form that may be deployed on the assembly platform. The architecture defines a framework for the exploitation of service oriented computing, web technologies, Internet of Thinks, concepts of Cyber Physical Systems, Big Data and the Cloud. It provides the infrastructure to address several challenges in the domain of flexible and evolvable hybrid Assembly Systems.

In our first collaboration with Mechanical engineers 6 years ago we realized the great communication gap that exists between the two domains, i.e., Electrical Eng and Mechanical Eng. Perhaps the biggest source of the communication gap is the use of terms with different semantics. The establishment of a basic terminology with well defined semantics is the first prerequisite to a productive and effective collaboration. Since this work is a specialization of the generic framework defined for Mechatronic industrial automation systems in the assembly systems domain, and we are not experts in this domain our next step is to further discuss this work with domain experts and extend and refine it to successfully address the demands of today's assembly industry.

References

[1] Hu S.J., Ko J., Weyand L., ElMaraghy H.A., Lien T.K., Koren Y., Bley H., Chryssolouris G., Nasr N., Shpitalni M., "*Assembly system design and operations for product variety*", CIRP Annals - Manufacturing Technology, Volume 60, pages 715-733, 2011
[2] Alsterman, H., Barata, J., and Onori, M. (2004). "Evolvable Assembly Systems Platforms: Opportunities and Requirements." Intelligent Manipulation and Grasping, R. Molfino, ed., IMG'2004, Genova, 18-23.
[3] M. Onori, J. Barata, R. Frey, "Evolvable assembly systems basic principles", W. Shen (Ed.), IT for Balanced Manufacturing Systems 220, IFIP, Springer, Boston (2006), pp. 317–328.
[4] Regina Frei, Luis Ribeiro, José Barata, Daniel Semere, "Evolvable Assembly Systems: Towards User Friendly Manufacturing", Proceedings of the 2007 IEEE International Symposium on Assembly and Manufacturing, Ann Arbor, Michigan, USA, July 22-25, 2007.
[5] OMG, "OMG Systems Modeling Language (OMG SysML™)," Version 1.3 June 2012.
[6] OMG, "Unified modeling language: Superstructure," *Version 2.4.1 2011*
[7] Craig Schlenoff, Amy Knutilla, Steven Ray "Unified Process Specification Language: Requirements for Modeling Process", NIST, National Institute of Standards and Technology, Manufacturing Engineering Laboratory, Manufacturing Systems Integration Division, Gaithersburg, MD 20899.
[8] N. Lohse, H. Hirani, S. Ratchev, M. Turitto, "An ontology for the definition and validation of assembly processes for evolvable assembly systems", ISATP, Proceedings of the IEEE International Symposium (2005), pp. 242–247.
[9] Dong Liang, Li Yuan, Yu Jian-feng, Zhang Jie, "A Cooperative Method between Assembly Process Planning and Simulation for Complex Product" International Conference on Interoperability for Enterprise Software and Applications, China, 2009.
[10] Gruninger, M., 2003. Ontology of the process specification language. In: Handbook of ontologies.
[11] Thramboulidis, K. "Model Integrated Mechatronics – Towards a new Paradigm in the Development of Manufacturing Systems" IEEE Transactions on Industrial Informatics, vol. 1, No. 1. February 2005.
[12] K. Thramboulidis, "The 3+1 SysML View-Model in Model Integrated Mechatronics", Journal of Software Engineering and Applications (JSEA), vol.3, no.2, 2010, pp.109-118.
[13] K. C. Thramboulidis, G. Doukas, and G. Koumoutsos, "A SOA-Based Embedded Systems Development Environment for Industrial Automation", EURASIP Journal on Embedded Systems Volume 2008, Article ID 312671, 15 pages, doi:10.1155/2008/312671
[14] Schmidt, D.C. (February 2006). "Model-Driven Engineering". *IEEE Computer* **39** (2).
[15] M. Bichler and K.J. Lin, "Service-oriented computing," *Computer*, vol. 39, no. 3, pp. 99–101, 2006.
[16] J. L. M. Lastra and M. Delamer, "Semantic web services in factory automation: fundamental insights and research roadmap," *IEEE Transactions on Industrial Informatics*, vol. 2, no. 1, pp. 1–11, 2006.
[17] O. Kaykova, O. Khriyenko, A. Naumenko, V. Terziyan, and A. Zharko, "RSCDF: a dynamic and context-sensitive metadata description framework for industrial resources," *Eastern-European Journal of Enterprise Technologies*, vol. 3, no. 3, pp. 55–78, 2005.
[18] Makris S., Pintzos G., Rentzos L., Chryssolouris G., "*Assembly support using AR technology based on automatic sequence generation*", CIRP Annals - Manufacturing Technology, Volume 62, pages 9-12, 2013
[19] Papakostas N., Michalos G., Makris S., Zouzias D., Chryssolouris G., "*Industrial applications with cooperating robots for the flexible assembly*", CIRP Annals - Manufacturing Technology, Volume 24, No. 7, pages 650-660, 2011
[20] K. Thramboulidis, "A Framework for the Implementation of Industrial Automation Systems Based on PLCs", (under review)
[21] Browne, J., Dubois, D., Rathmill, K., Sethi, S.P., Stecke, K.E., "Classification of Flexible manufacturing systems." The FMS Magazine 2 (2), 1984, pp.114-117.
[22] R. Beach, A.P. Muhlemann, D.H.R. Price, A. Paterson, J.A. Sharp, "A review of manufacturing flexibility", European Journal of Operational Research, 122 (2000), pp. 41–57.
[23] Hoda A. ElMaraghy, "Flexible and reconfigurable manufacturing systems paradigms", International Journal of Flexible Manufacturing Systems, October 2005, Volume 17, Issue 4, pp 261-276
[24] S. Burmester, H. Giese, and M. Tichy, "Model-driven development of reconfigurable mechatronic systems with mechatronic 'UML' in model driven architecture," Springer Berlin/Heidelberg, Vol. 3599, 2005.
[25] INCOSE Systems Engineering Vision 2020 INCOSE-TP-2004-004-02 September, 2007.
[26] Yue Cao, Yusheng Liu, Christiaan J.J. Paredis, "System-level model integration of design and simulation for mechatronic systems based on SysML", Mechatronics 21 (2011) 1063–1075.
[27] Niels Lohse, Hitendra Hirani, Svetan Ratchev "Equipment ontology for modular reconfigurable assembly systems", International Journal of Flexible Manufacturing Systems, October 2005, Volume 17, Issue 4, pp 301-314
[28] Gonçalo Cândido, Armando W. Colombo, José Barata, François Jammes, "Service-Oriented Infrastructure to Support the Deployment of Evolvable Production Systems", IEEE transactions on Industrial Informatics, Vol. 7, No. 4, Nov 2011.
[29] Makris S., Michalos G., Eytan A., Chryssolouris G., "*Cooperating robots for reconfigurable assembly operations: Review and challenges*", (CMS2012), 45th CIRP Conference on Manufacturing Systems, Athens, Greece, pp.393-399, 2012
[30] Xu HM, Yuan MH, Li DB, "*A Novel Process Planning Schema Based on Process Knowledge Customization.*" The International Journal of Advanced Manufacturing Technology 44(1), 2009, pp.161–172.
[31] AlGeddawy T, ElMaraghy HA, "Design of single assembly line for the delayed differentiation of product variants." Flexible Services and Manufacturing Journal 22(3–4), 2010, pp. 163–182.
[32] Hu SJ, Stecke KE, "Analysis of Automotive Body Assembly System Configurations for Quality and Productivity.", International Journal of Manufacturing Research 4, 2009, pp.117–141.